\documentclass[twocolumn,showpacs,preprintnumbers,amsmath,amssymb]{revtex4}

\usepackage{graphicx}
\usepackage{dcolumn}
\usepackage{bm}

\begin{document}
\date{\today}
\title{Ultrafast response of surface electromagnetic waves in an aluminum film perforated with subwavelength hole arrays}
\author{M. Tong,$^{1}$ A. S. Kirakosyan,$^{2,3}$ T. V. Shahbazyan,$^2$ and  Z. V. Vardeny$^1$}

\affiliation{$^1$Department of Physics, University of Utah, Salt Lake
  City, Utah 84112 USA\\
$^2$Department of Physics, Jackson State University, Jackson, MS
39217 USA\\
$^3$Department of Physics, Yerevan State University, 
Yerevan, 375025 Armenia
}

\begin{abstract}
The ultrafast dynamics of surface electromagnetic waves photogenerated on aluminum film perforated with subwavelength holes array was studied in the visible spectral range by the technique of transient photomodulation with $\sim 100$ fs time resolution. We observed a pronounced \emph{blueshift} of the resonant transmission band that reveals the important role of plasma attenuation in the optical response of nanohole arrays. The blueshift is inconsistent with plasmonic mechanism of extraordinary transmission and points to the crucial role of interference in the formation of transmission bands. The transient photomodulation spectra were successfully modeled within the Boltzmann equation approach for the electron-phonon relaxation dynamics, involving non-equilibrium hot electrons and quasi-equilibrium phonons.
\end{abstract}


\maketitle

Light transmission through periodic subwavelength arrays of holes in optically thick metal films \cite{ebbesen98} (plasmonic lattices) has attracted intensive interest due to potential applications in near-field microscopy, optoelectronics, and biosensing  \cite{barnes03,rigneault05}. On a smooth metal-dielectric interface light cannot couple to surface electromagnetic waves (SEW) such as, e.g., surface plasmon polaritons (SPP) because energy and momentum conservation cannot be simultaneously obeyed. But the periodicity in nanohole arrays leads to the formation of band structure that enables light to directly excite SEW, which consequently mediate the optical transmission through the metal film. The transmission spectrum, $T_0(\omega)$, is then characterized by several resonant bands, each followed by a minimum caused by Wood's anomaly that is due to the diffraction grating related to the corrugated film \cite{ebbesen98}. These “extraordinary optical transmission” (EOT) resonances were originally attributed to grating coupling of light to SPP excitations on both sides of the perforated film \cite{ebbesen98,ghaemi98,ebbesen01}; but later other possible explanations were also suggested such as interference of resonant cavity modes  \cite{porto99,lalanne00} and dynamical diffraction \cite{treacy02,lezec04}. The EOT spectrum has been shown to depend sensitively on geometric parameters such as film thickness, lattice periodicity $a$, and holes size and shape \cite{koerkamp04}; as well as on the material optical constants such as substrate ($\varepsilon_d$) and hole ($\varepsilon_h$) dielectric constants, and that of the metal film: $\varepsilon_{m}(\omega) = \varepsilon_{m}'+ i \varepsilon_{m}''$. However, relatively little is known about the EOT dynamics that arises when the optical constants are subjected to a \textit{pulsed time-dependent perturbation}. It has long been known \cite{chemla01} that time-resolved ultrafast spectroscopy can provide valuable insight into the makeup of underlying excitations from the transient changes of optical spectra \cite{shahbazyan01}.

Here we investigated the \textit{ultrafast dynamics} of the EOT spectrum in Al-based plasmonic lattice measured by the transient photomodulation (PM) spectroscopy in the visible range with $\sim 100$ fs resolution. Surprisingly, we found a transient \textit{blueshift} of the main EOT band, which unravels the important role of the plasma \textit{attenuation} in determining the dynamical response of nanohole arrays. The blueshift is inconsistent with the simple SPP mechanism of EOT, and indicates the dominant role of interference in the formation of the transmission bands. The obtained transient dynamics is modeled within the Boltzmann equation for electron-phonon relaxation with non-equilibrium hot electrons and quasi-equilibrium phonons, and subsequent heat transfer to the glass substrate.

The plasmonic lattice sample was a 70 nm thick Al film, 5 $\times$ 5 mm$^2$ in area deposited on a glass substrate. The metal film was perforated with circular holes having diameter $D = 150$ nm in a square hole array with lattice constant  $a = 300$ nm resulting in a fractional aperture area of 25\%. The ultrafast laser system used for measuring the transient PM spectrum was a Ti:sapphire regenerative amplifier with pulses of 100 fs duration at photon energies of 1.55 eV, having 400 $\mu$J energy per pulse at a repetition rate of 1 kHz. The second harmonic of the fundamental pulses at 390 nm (3.1 eV) were used as the pump beam. The probe beam was a white light super-continuum generated in a 1-mm thick sapphire plate that covers the spectral range from 1.6 to 2.8 eV; the white light super-continuum was also used for measuring the unperturbed transmission,  $T_0(\omega)$. The pump and probe beams were directed onto the perforated Al film from the air side to maximize coupling of light to the SEW. The pump and probe beams were directed onto the perforated Al film from the air side to maximize coupling of light to the SEW. The transient PM spectrum was obtained from the photoinduced change ($\Delta T$) in $T_0$ using a phase-sensitive technique with a resolution $\Delta T/T_0\sim 10^{-4}$  that corresponds to a photoexcited electron density in the Al metal of $\sim 10^{17}$ cm$^{-3}/$pulse.  

\begin{figure}
\begin{center} 
\includegraphics[width=0.85\columnwidth]{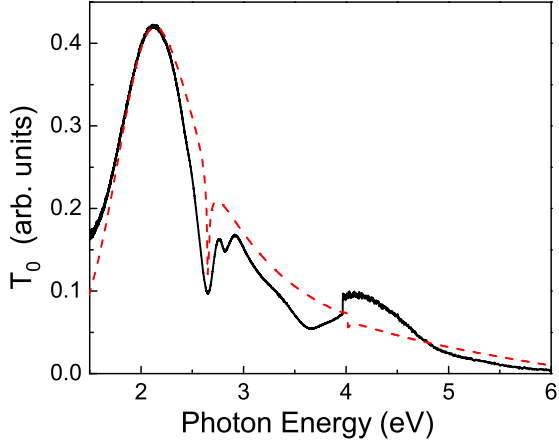}
\caption{
(Color online) The measured (solid line) and calculated (dashed line) normal incidence optical transmission, $T_0$ through an Al plasmonic lattice comprised of a square array of nanoholes with lattice constant 300 nm and hole diameter 150 nm. 
} 
\label{fig:1}
\end{center}
\end{figure}

Figure \ref{fig:1} shows the EOT spectrum, $T_{0}(\omega)$, of the Al plasmonic lattice at normal incidence. Several EOT bands and transmission anti-resonances (AR) are observed.  We note that  $T_0(\omega)$ at resonance frequencies substantially surpasses the fractional aperture area of 25\%. The lowest frequency EOT band peaks at $\sim 2.1$ eV; whereas the lowest frequency AR feature that is associated with Wood's anomaly at the Al/glass interface with wavevector  $k^{(01)} = g \varepsilon_{gl}^{-1/2}$ (where $\varepsilon_{gl}=2.43$  and $g=2\pi/a$ ) is in the form of a dip at 2.6 eV.  The AR's at higher frequencies are higher-order Wood's anomalies that correspond to wavevectors $k^{(mn)}= g\left( m^2 + n^2\right)^{1/2}\varepsilon_d^{-1/2}$ (\textit{m, n} are integers) as follows:  3.7 eV ($k^{(11)}  = g \varepsilon_{gl}^{-1/2} 2^{1/2}$) that corresponds to the (1,1) branch at the Al/glass interface; and 4.0 eV ($k^{(01)} = g$) that corresponds to the (0,1) branch at the Al/air interface.  It is noteworthy that the lowest frequency EOT resonance (at $\sim 2.1$ eV) occurs at considerably lower (by $\sim 0.4$ eV) frequency than that predicted by the lowest-order SPP having wavevector that satisfies the Bragg condition \cite{ebbesen98}: $k_{spp}^{(mn)}  = g \left[ \left( m^2 + n^2\right) \left( \varepsilon_{m}'^{-1} + \varepsilon_{d}^{-1} \right) \right]^{1/2}$. 

\begin{figure}
\centering
\includegraphics[width=0.85\columnwidth]{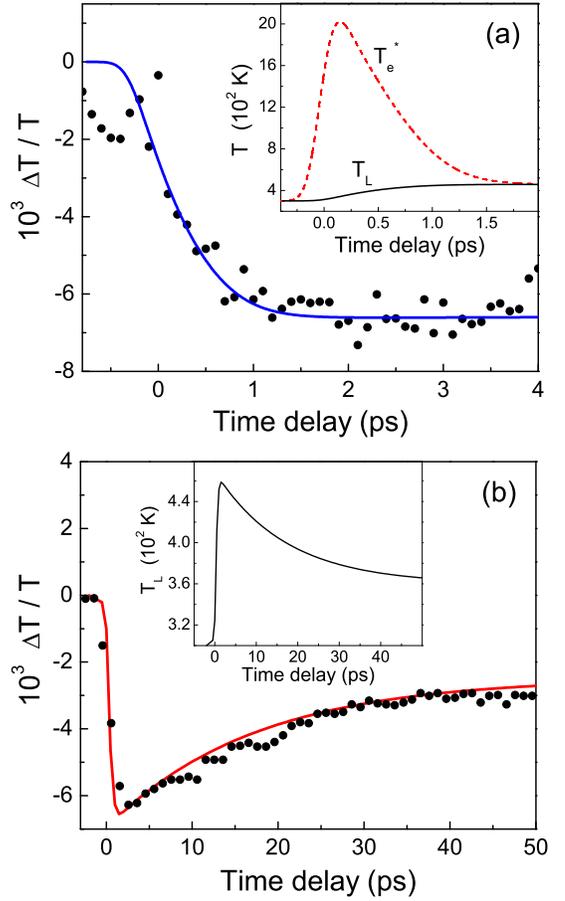}
\caption{ 
(Color online) Short-time (a) and long-time (b) dynamics of the transient PM of the Al plasmonic lattice at 1.9 eV.  The solid line is a fit based on Eqs.~(\ref{t00}-\ref{2tphonon}) in the text. Insets: (a) short-time evolution of the Al effective electron temperature, $T_e^*$ (solid line) and lattice, $T_L$ (dashed line); (b) long-time evolution of $T_L$.
} 
\label{fig:2}
\end{figure}

Figure \ref{fig:2} shows the PM transient response (photobleaching) at 650 nm (1.9 eV). Two distinct dynamic responses are apparent: (i) a PM rise during $\sim 1$ ps followed by a \textit{plateau} [Fig.~\ref{fig:2}(a)]; (ii) after about 3 ps the signal starts recovering with characteristic time constant of $\sim 40$ ps [Fig.~\ref{fig:2}(b)]. These are unique features associated with the Al plasmonic lattice. In contrast to noble metals \cite{fatti00}, the interband transitions in Al do not play a significant role in determining  $\varepsilon_{m}(\omega)$, and therefore the PM response dynamics stems from the transient change in the electron-phonon (\textit{e-p}) relaxation rate, $\gamma_{ep}(T_L)$, where $T_L$ is the metal lattice temperature.  The dielectric function $\varepsilon_{m}(\omega)$ of Al is usually approximated by a Drude form; $\varepsilon_{m} = 1 - \omega_p^2/\omega (\omega+ i \gamma)$, where $\omega_p$  is the bulk plasma frequency, and  $\gamma = \gamma_0 +  \gamma_{ep}$  is the attenuation with relatively small contribution,  $\gamma_0$, from electron-electron (\textit{e-e}) and impurity scattering \cite{smith82}. An increase in  $T_L$ that is caused by energy exchange between the photoexcited (hot) electrons and phonons, is accompanied by an increase in  $\gamma_{ep}(T_L)$ due to larger phase space available for the phonon states at elevated temperature. The transient change in  $\varepsilon_{m}(\omega)$ leads to the shift of EOT resonant frequencies (but less so for the AR frequencies, which are mainly determined by the plasmonic lattice periodicity). We thus interpret the PM rise of $\sim 1$ ps as due to the increase in $\gamma$ (and consequently also in $\varepsilon_{m}''$) caused by \textit{e-p} energy exchange. Furthermore, due to the strong \textit{e-p} interaction in Al, the excess energy transfer to the lattice occurs \textit{before} the electron thermalization process is completed; therefore the short-time dynamics is due to an inherently non-equilibrium process. After an equilibrium between the electron and phonon subsystems in the Al film is reached at higher $T_L$ [plateau in Fig.~\ref{fig:2}(a)], the excess heat dissipates into the glass substrate within $\sim 40$ ps [Fig.~\ref{fig:2}(b)].

\begin{figure}
\begin{center} 
\includegraphics[width=0.85\columnwidth]{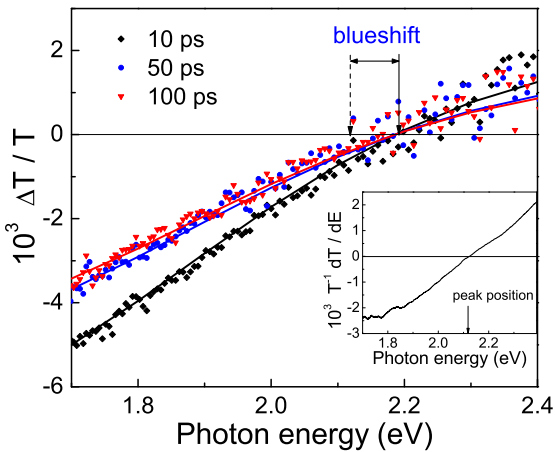}
\caption{
(Color online) Transient PM spectra at various time delays; arrows indicate the blueshift. The solid lines are fits based on Eqs.~(\ref{t00}-\ref{2tphonon}) in the text. Inset: first derivative of $T_0$ spectrum; the arrow points to the frequency of the EOT band maximum in $T_0$ spectrum.
} 
\label{fig:3}
\end{center}
\end{figure}

Figure \ref{fig:3} shows several transient PM spectra at various times, $t$ following the pulsed excitation, in the frequency range that roughly covers the main EOT band. Although the data becomes noisier close to the lowest Wood's anomaly, a clear deviation from linear behavior emerges that cannot be reproduced by a simple first derivative of the EOT band (Fig.~\ref{fig:3} inset); this indicates that for plasmonic lattices the various spectral features in $T_{0}(\omega)$ have distinctively different response to a pulsed excitation. A noticeable \textit{blueshift} ($\sim 70$ meV) of the main EOT band is apparent that persists for times up to 100 ps. The blueshift indicates that the EOT is sensitive to the imaginary part of permittivity, $\varepsilon_{m}''(\omega)$, for the following reason. Within the Drude model for Al, we have  $\varepsilon_{m}'(\omega) = 1 - \omega_{p}^2/(\omega^2 + \gamma^2)$. With such $\varepsilon_{m}'$ and for $\omega\ll \omega_{p}$, the above-mentioned SPP Bragg condition leads to the following expression for, e.g., the (0,1) mode frequency:
\begin{eqnarray}
\label{spp-bragg}
\omega_{spp}^{01}\approx \omega_{g}\left[ \varepsilon_{d}^{-1}-(\omega_{g}^{2}+\gamma^{2})/\omega_{p}^{2}\right]^{1/2}
\end{eqnarray}
where $\omega_{g}=cg$ is the frequency associated with the reciprocal lattice period ($c$ is the speed of light). Consequently, if SPP is the cause of the resonant transmission, a transient increase in  $\gamma$ should lead to  a weak \emph{redshift} of the resonance frequency, in \emph{disagreement} with the data. We thus conclude that the observed spectral blueshift cannot be explained by the present version of the SPP mechanism for the EOT phenomenon \cite{ebbesen98}. At the same time, an increase in $\gamma$  causes an increase in the \textit{attenuation},  $\varepsilon_{m}''$, which for $\gamma\ll\omega$ can be approximated by  $\varepsilon_{m}''  \approx \gamma \omega_{p}^{2}/\omega^3$. If   $\varepsilon_{m}''$ participates in determining the position and lineshape of the EOT bands, then the increase in  $\gamma$ implies \textit{higher} resonant frequencies (blueshift), in \emph{agreement} with the data. 

For incorporating the non-trivial role of attenuation in determining the EOT resonant bands, we adopt an analytical effective-medium model \cite{kirakosyan2bp} that is a combination of the \textit{one-mode} model \cite{lalanne00} and \textit{weakly modulated} permittivity model \cite{darmanyan03}; these are not directly related to SPP excitations. In our model, transmission through the film occurs via coupling of the incident light to the eigenmodes of the periodic system \cite{treacy02}. Consequently,  $T_0(\omega)$ at normal incidence is given by
\begin{equation}
\label{t00}
T_0= \Biggl|\frac{t_{1 2} t_{2 3} e^{- \kappa h} }{1 + r_{1 2} r_{2 3} e^{- 2 \kappa  h} }\Biggr|^2,
\end{equation}
where $h$ is the film thickness, $t_{12} = 2 \tilde{n}_{2} /( \tilde{n}_{1} + \tilde{n}_{2} )$ and $t_{23} = 2 n_{3} /( \tilde{n}_{2} + \tilde{n}_{3} )$ are  transmission amplitudes through air/metal and metal/glass interfaces, and $r_{k l} = (\tilde{n}_{k} - \tilde{n}_{l} )/( \tilde{n}_{k} + \tilde{n}_{l} )$ are the corresponding reflection amplitudes, $\kappa$ is the fundamental mode eiegenvalue, and $\tilde{n}_{k}(\omega)$ are \textit{effective} refraction indices of each media, which are determined by overlap integrals between electromagnetic eigenmodes of the perforated film and dielectric media \cite{lalanne00}. The overlap integrals, which are defined at the metal/dielectric interfaces, are evaluated using weak modulation approximation \cite{darmanyan03}, whereas the eigenvalue  $\kappa$, which is determined by the bulk part of the perforated metal, is calculated by using the analogy that exists between Maxwell and Schr\"{o}dinger equations within a complex Kr\"{o}nig-Penney potential \cite{kirakosyan2bp}. The fit to $T_{0}(\omega)$  using this model with appropriate parameters taken directly from the experimental conditions  ($\omega_p = 15$ eV and  $\gamma =  0.2$ eV for Al) is shown in Fig.~\ref{fig:1}. The main EOT band and first Wood's anomaly in $T_0(\omega)$ are accurately described by our model; at higher frequencies, where second order features are apparent, the fit is less accurate due to the one-mode approximation used in the model.

To model the ultrafast dynamics at low photoexcited carrier density, we calculated the changes in  $T_0$ [Eq.~(\ref{t00})], by replacement $\varepsilon_{m}(\omega) \rightarrow \varepsilon_{m}(\omega) + \delta\varepsilon_{m}(\omega, t)$, where  
\begin{equation}
\label{deltaepsilon}
 \delta \varepsilon_{m}(\omega, t) = i \omega_p^2 
\frac{\gamma\left[T_L(t)\right] - \gamma_0}
{\left[ \omega (\omega + i \gamma_0)\right]^2 } 
\end{equation}
is the photomodulated Drude permittivity.  The \textit{e-p} contribution to the relaxation rate in Eq.~(\ref{deltaepsilon}) is given by  $\gamma_{e p}(T_L) = B (T_L/\Theta_D)^5 J(\Theta_D/T_L)$, where $J(y) = \int_{0}^{y} dx x^4 \coth(x/2)/2$,  $\Theta_D\approx 380$ K  is the Debye temperature for Al, and B is a material-dependent constant \cite{gotze&wolfle72}. Importantly, due to the strong \textit{e-p} interaction in Al, the energy exchange between the hot electrons and phonons occurs simultaneously with the thermalization of non-equilibrium electron population via \textit{e-e} scattering. This causes the short-time dynamics in Al to be an inherently non-equilibrium process. However, since the phonons gain energy only from the hot electrons, their distribution can be considered as quasi-equilibrium  \cite{fatti00,grua03}, and thus characterized by a time-dependent temperature $T_L(t)$ that satisfies the rate equation \cite{kirakosyan2bp}
\begin{equation}
C_L \frac{\partial T_L}{\partial t} = G (T^{*}_e - T_L) +S_{GL}(t),
\label{2tphonon}
\end{equation}
where $C_L= 900 $ Jkg$^{-1}$K$^{-1}$ is the Al lattice heat capacity, $G = 3.1 \times
10^{17}$ Wm$^{-3}$K$^{-1}$ is the \emph{e-p} interaction constant, and  $T_{e}^* = \int dE f(E, t) \left[1 - f(E, t) \right] $ plays the role of effective temperature for the \textit{athermal} non-equilibrium electrons having a distribution function $f(E, t)$ (which for thermal distribution coincides with the usual hot electron temperature). The last term in Eq.~(\ref{2tphonon}) describes the heat exchange between the Al film and glass substrate; it is given by $S_{GL}(t)=\kappa_G\, \partial T_G/\partial z$, where $T_G$ and $\kappa_G=1.37$ Jm$^{-1}$K$^{-1}$ are the glass temperature and thermal conductivity, respectively. The rate equation  (\ref{2tphonon}) is supplemented by heat transport equation in the glass that governs the long-time dynamics \cite{grua03}. In contrast, the short-time dynamics is dominated by the energy transfer from the non-equilibrium electron population to the lattice, and was simulated here by a full numerical solution of the Boltzmann equation for $f(E, t)$ \cite{kirakosyan2bp}.

The results of our numerical simulations are shown in Figs.~\ref{fig:2} and \ref{fig:3}. The calculated dynamics are in excellent agreement with the data, including the PM rise with sub-picosecond time constant [Fig.~\ref{fig:2}(a)] as well as the long-time decay [Fig.~\ref{fig:2}(b)]. The two insets in  Fig.~\ref{fig:2} show evolutions of the calculated lattice and effective electron temperatures. PM dynamics can be summarized as follows. After photoexcitation, the effective electron temperature $T_{e}^*$ shows a sharp rise that indicates thermalization of the electron gas via \textit{e-e} scattering. The subsequent drop of $T_{e}^*$ for $t\leq 1$ ps after reaching its peak value at $\sim 200$ fs, which is accompanied by an increase in $T_L$ and hence in  $\Delta T/T_0$, indicates energy intake by the lattice. Note that the energy transfer to the lattice takes place while the electron gas is still in the thermalization stage, in contrast to the dynamics in noble metals \cite{fatti00,grua03}. This is confirmed by numerical calculation of $f(E, t)$ (not shown) indicating the completion of electron thermalization by $t_T < 1$ ps \cite{kirakosyan2bp}. As the electron and phonon subsystems equilibrate at $t_T$, PM reaches a plateau persisting until the slower process of heat transfer to the substrate takes over [Fig.~\ref{fig:2}(b)].

Fits of the transient PM spectra based on our model are shown in Fig.~\ref{fig:3} for $t>t_T$. The fits capture the main EOT band blueshift as well as the nonlinear behavior of the PM spectrum close to the Wood’s anomaly. The nonlinearity stems from the fact that the transmission spectrum in plasmonic lattices does not respond uniformly to an increase in the attenuation. This happens since the AR frequencies are determined by the \emph{lattice periodicity}, and thus, in contrast to the resonant bands, they do not shift when $\gamma$ changes. The PM blueshift can be traced back to $\Delta T$ dependence on $\varepsilon_{m}''$, as discussed above. The eigenmodes in a plasmonic lattice are the result of interference of multiple waves from the perforated metal/dielectric interfaces. In the presence of attenuation, the scattered wave phase has a contribution $\propto \mbox{Re} \sqrt{\varepsilon_{m}} = 	\varepsilon_{m}''/\sqrt{|\varepsilon_{m}'|} \approx \gamma \omega_p/\omega^2$. Therefore, a transient change in attenuation causes corresponding shifts in the scattering phases. This, in turn, changes the interference of waves that form an eigenmode. To compensate for the phase shifts, a transient increase in $\gamma$ must be accompanied by a corresponding increase in $\omega$; and consequently the transmission resonances blueshift.

In summary, we investigated the ultrafast dynamics of nanohole array on Al film that reveals an important role played by the plasma attenuation in the makeup of surface electromagnetic modes. The transient optical spectra discriminate between electromagnetic excitations with different response to changes in the metal dielectric function. The observed transient blueshift favors the interference-based mechanisms of explaining the EOT phenomenon, as opposed to SPP excitation as the cause for the resonant transmission bands.

This work was supported in part by the NSF Grant No. DMR-0503172, and the SYNERGY program at the University of Utah. Work at JSU was supported in part by the NSF under Grant No. DMR-0606509 and EPSCOR program, and the NIH Grant No. 2 S06 GM008047-33.

\end{document}